\documentclass[aps,preprintnumbers,superscriptaddress,showpacs]{revtex4}
\usepackage{epsfig}
\usepackage{psfrag}
\usepackage{amsfonts}
\usepackage{graphicx}
\usepackage{dcolumn}
\usepackage{bm}

\begin{document}

\title{Light quark energy loss in a soft-wall AdS/QCD model}

\author{Xiangrong Zhu}
\email{xrongzhu@zjhu.edu.cn} \affiliation{School of Science,
Huzhou University, Huzhou 313000, China}

\author{Zi-qiang Zhang}
\email{zhangzq@cug.edu.cn} \affiliation{School of Mathematics and
Physics, China University of Geosciences, Wuhan 430074, China}

%%%%%%%%%%%%%%%%%%%%%%%%%%%%%%%%%%%%%%%%
\begin{abstract}
We investigate the energy loss of light quarks in a holographic
QCD model with conformal invariance broken by a background
dilaton. We perform the analysis within falling string and
shooting string, respectively. It turns out that the two methods
give the same result: the presence of chemical potential and
confining scale tends to enhance the energy loss, in accord with
previous findings of drag force and jet quenching parameter.

\end{abstract}
\pacs{11.25.Tq, 11.15.Tk, 11.25-w} \maketitle
%%%%%%%%%%%%%%%%%%%%%%%%%%%%%%%%%%%%%%%%
\section{Introduction}
One of the main purposes of the heavy-ion collisions experiments
is to explore the QCD phase diagram and the properties of new
state of matter produced through collisions at high energy
density. It is believed that the experimental program at
Relativistic Heavy-Ion Collider (RHIC) and Large Hadron Collider
(LHC) have created a new state of matter so-called quark gluon
plasma (QGP) \cite{ev,mg}. One of the striking features of such
substance is jet quenching: the energy loss of high energy partons
produced through collisions as they interact with the plasma
before they fragment into hadrons (for recent reviews see
\cite{mc,gy}). On the other hand, it has been observed that QGP
behaves as a strongly coupled fluid \cite{es,kad,ja}, which
involves nonperturbative physics suitable for the application of
the anti-de Sitter/conformal field theory (AdS/CFT)
correspondence.

AdS/CFT \cite{Maldacena:1997re,Gubser:1998bc,MadalcenaReview}, the
duality between the type IIB superstring theory formulated on
AdS$_5\times S^5$ and $\mathcal N=4$ super Yang-Mills theory (SYM)
in four dimensions, provides a useful tool for understanding the
strong interaction, i.e. quantum chromodynamics (QCD). Although
SYM differs from QCD in many properties at zero temperature, it
reveals some qualitative features of QCD in the strongly coupled
regime at non-zero temperature. During the last two decades, the
AdS/CFT correspondence has yielded many important insights for
studying various aspects of QGP (see \cite{JCA,OD} for recent
reviews with many phenomenological applications). An interesting
example of such applications is jet quenching. For instance, the
drag force which describes the energy loss for heavy quarks moving
through $\mathcal N=4$ SYM plasma was studied in \cite{cp,ss0}.
Moreover, the jet quenching of light quarks moving through
$\mathcal N=4$ SYM plasma has been addressed based on various
approaches, e.g., jet quenching parameter \cite{hl,hl1}, falling
string \cite{ss,pm,pm1,pa,pa1}, shooting string \cite{ss1,ss2},
etc.

Here we present an alternative holographic approach to study the
energy loss of light quarks. The motivation is that AdS/QCD
models, e.g., hard wall \cite{H1,H2}, soft wall \cite{AKE} and
some improved holographic models \cite{JP,AST,DL,DL1,SH,SH1,RRO}
could provide a nice phenomenological description of
quark-antiquark interaction and some hadronic properties. In
particular, we will employ the SW$_{T,\mu}$ model \cite{PCO} which
is defined by the AdS with a charged black hole to describe finite
temperature and density multiplied by a warp factor to generate
confinement. It turns out that this model has achieved
considerable success in describing various aspects of hadron
physics \cite{PCO,CPA,PCO1,PCO2,YH,zq0,XCH,zq}. Another motivation
for this paper is that the drag force and jet quenching parameter
have been studied in the SW$_{T,\mu}$ model \cite{YH,zq0} and the
results show that the presence of chemical potential and confining
scale increases the two parameters thus enhancing the energy loss.
Inspired by this, we wonder whether chemical potential and
confining scale have the same effect on the energy loss of light
quarks using falling string and shooting string.

The outline of the paper is as follows. In the next section, we
introduce the SW$_{T,\mu}$ model presented in \cite{PCO}. In
section 3, we study the energy loss of light quarks in this model
within falling string and shooting string, in turn. The last part
is devoted to conclusion and discussion.

\section{Setup}
In this section, we briefly review the SW$_{T,\mu}$ model given in
\cite{PCO}. In the string frame, the SW$_{T,\mu}$ model has the
following metric
\begin{equation}
ds^2=\frac{R^2}{z^2}h(z)(-f(z)dt^2+d\vec{x}^2+\frac{dz^2}{f(z)}),\label{metric}
\end{equation}
with
\begin{equation}
f(z)=1-(1+Q^2)(\frac{z}{z_h})^4+Q^2(\frac{z}{z_h})^6, \qquad
h(z)=e^{c^2z^2},
\end{equation}
where $R$ is the AdS radius. $Q$ is the charge of black hole. $z$
represents the fifth coordinate with $z=z_h$ the horizon, defined
by $f(z_h)=0$. The warp factor $h(z)$, characterizing the soft
wall model, distorts the background metric and brings the
confining scale $c$ (see \cite{JP} for a anatlytical way to
introduce the warp factor within potential reconstruction
approach).

The temperature of the black hole reads
\begin{equation}
T=\frac{1}{\pi z_h}(1-\frac{Q^2}{2}), \qquad 0\leq
Q\leq\sqrt{2}.\label{T}
\end{equation}

The chemical potential reads
\begin{equation}
\mu=\sqrt{3}Q/z_h.
\end{equation}

Notice that for $Q=0$, the SW$_{T,\mu}$ model becomes the Andreev
model \cite{OA}. For $c=0$, it reduces to the AdS-Reissner
Nordstrom black hole \cite{CV:1999,DT:2006}. For $Q=c=0$, it
restores to the AdS black hole.
\section{Energy loss of light quarks in the SW$_{T,\mu}$ model}

\subsection{Falling string}
In previous research, various authors \cite{ss,pm,pm1,pa,pa1} have
studied the jet quenching of light quark in $\mathcal N=4$ SYM
plasma using falling string in different ways, e.g, with or
without the addition of fundamental-charge matter, specify
different initial conditions. Though each approaches the subject
from different vantage points, the analysis of the stopping
distance traveled by the falling string in $AdS_5$-Schwarzschild
are consistent.

Next, we will follow the argument in \cite{pa,pa1} to study the
jet quenching of light quark in the SW$_{T,\mu}$ model by
analyzing the stopping distance of an image jet induced by a
massless source field, characterized by a massless particle
falling along the null geodesic in the WKB approximation.
According to this scenario, the R-charged current is generated by
a massless gauge field in the gravity dual and the induced current
is regarded as an energetic jet passing through the medium. When
the wave packet of the massless gauge field falls into the horizon
of the dual geometry, the image jet on the boundary dissipates and
then thermalizes in the medium. The stopping distance is therefore
defined as the (maximum) distance for a jet passing through the
medium before it thermalizes.

In the WKB approximation, the wave packet of the massless gauge
field in the gravity dual is supposed to be localized in the
momentum space such that the wave function of the gauge field
could be factorized as
\begin{equation}
A_j(t,z)=exp[\frac{i}{\hbar}(q_kx_k+\int dzq_z)]\tilde{A}_j(t,z),
\end{equation}
where $q_k$ denotes the 4-momentum, conserved as the metric
preserves the translational symmetry along the 4-dimensional
spacetime. $q_z$ represents the momentum along the bulk direction.
$\tilde{A}_j(t,z)$ refers to the slow-varying with respect to $t$
and $z$. $j,k$ are the 4-dimensional spacetime coordinates.

In the classical limit, i.e., $\hbar\rightarrow 0$, the equation
of motion of the wave pack reduces to a null geodesic,
\begin{equation}
0=(ds^2)=dx^ig_{ij}dx^j+dzg_{zz}dz,
\end{equation}
yielding
\begin{equation}
\frac{dz}{d\zeta}=\frac{1}{\sqrt{g_{zz}}}[-g_{ij}\frac{dx^i}{d\zeta}\frac{dx^j}{d\zeta}]^{1/2},\label{n1}
\end{equation}
where $\zeta$ is an affine parameter for the trajectory. As the
4-dimensional translation invariance,
\begin{equation}
g_{ij}\frac{dx^j}{d\zeta},
\end{equation}
is conserved and proportional to $q_i$, leading to
\begin{equation}
\frac{dx^i}{d\zeta} \propto g^{ij}q_j.\label{n2}
\end{equation}

Then, dividing (\ref{n2}) by (\ref{n1}) gives
\begin{equation}
\frac{dx^i}{dz}=\sqrt{g_{zz}}\frac{g^{ij}q_j}{(-q_kg^{kl}q_l)^{1/2}},
\label{stop}
\end{equation}
one can check that the null geodesic in the above equation remains
unchanged even when one uses the Einstein frame.

To proceed, we calculate the stooping distance. Supposing that the
3-momentum $\vec{q}$ to point in one of $\vec{x}$ directions,
e.g., the $x_3$ direction, implying $q_i=(-\omega,0,0,|\vec{q}|)$,
where $\omega$ and $\vec{q}$ are the energy and spacial momentum
of the light quark, respectively. Then plugging (\ref{metric})
into (\ref{stop}), the stopping distance for the SW$_{T,\mu}$
model can be obtained,
\begin{equation}
x=\int_0^{z_h}\frac{dz}{e^{c^2z^2}\sqrt{\frac{\omega^2}{|\vec{q}|^2}
-\frac{1-(1+\frac{\mu^2z_h^2}{3})\frac{z^4}{z_h^4}+\frac{\mu^2z^6}{3z_h^4}}{e^{c^4z^4}}}},
\label{stop1}
\end{equation}
note that when one turns off the chemical potential and confining
scale effects by setting $\mu=c=0$, the above equation recovers
the result of SYM \cite{pa,pa1}.

Before going further, we need to turn to numerics. First, we
determine the value range of $c$. Here we tend to study the light
quark energy loss in a class of models parametrized by $c$ rather
than in a specific model with fixed $c$. To that end, we make $c$
dimensionless by normalizing it at fixed $T$ and express $\mu$ in
unit of it as well. The knowledge of lattice calculations suggests
\cite{HLL} that the range of $0\leq c/T\leq2.5$ is most relevant
for a comparison with QCD. We take that range here.

In fig.1, we compare the stopping distance of a light quark moving
in the SW$_{T,\mu}$ model with the same one moving in $\mathcal
N=4$ SYM plasma. The left panel corresponds to $x/x_{SYM}$ versus
$\mu/T$ with fixed $c/T$ while the right one represents
$x/x_{SYM}$ versus $c/T$ with fixed $\mu/T$, where $x_{SYM}$
denotes the stopping distance of SYM. From these figures, one
finds that the stopping distance in the SW$_{T,\mu}$ model is
smaller than that of SYM. In particular, the left panel tells us
that with fixed $c/T$, increasing $\mu/T$ leads to decreasing
$x/x_{SYM}$. Namely, the inclusion of chemical potential decreases
the stopping distance thus enhancing the energy loss, in accord
with that found in \cite{EL}. On the other hand, one can see from
the right panel that the confining scale has similar effect. Given
the above, one could reach the following conclusion: the presence
of chemical potential and confining scale both decrease the
stopping distance thus enhancing the energy loss, consistently
with the findings of the drag force \cite{YH} and jet quenching
parameter \cite{zq0}.

\begin{figure}
\centering
\includegraphics[width=8cm]{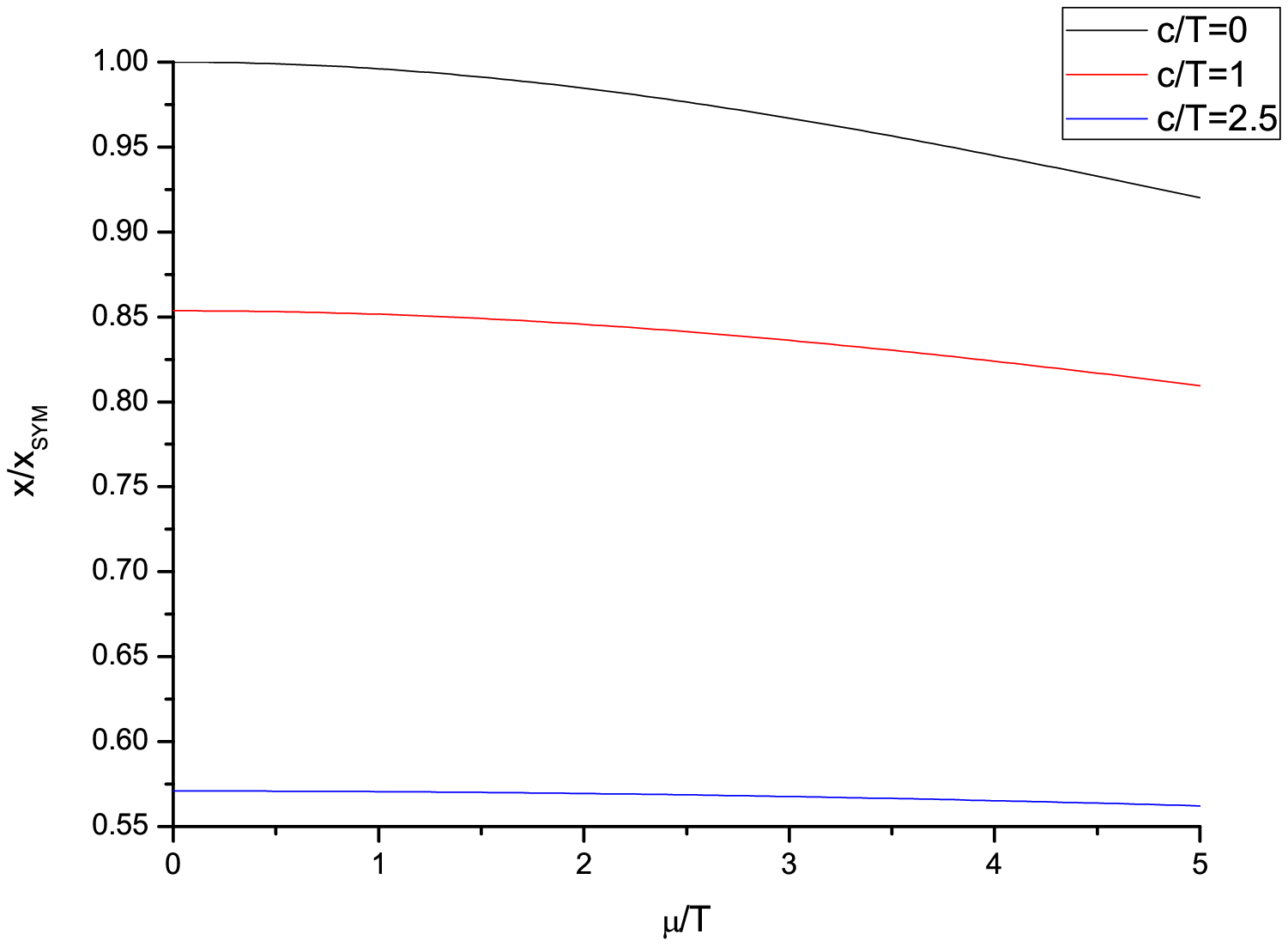}
\includegraphics[width=8cm]{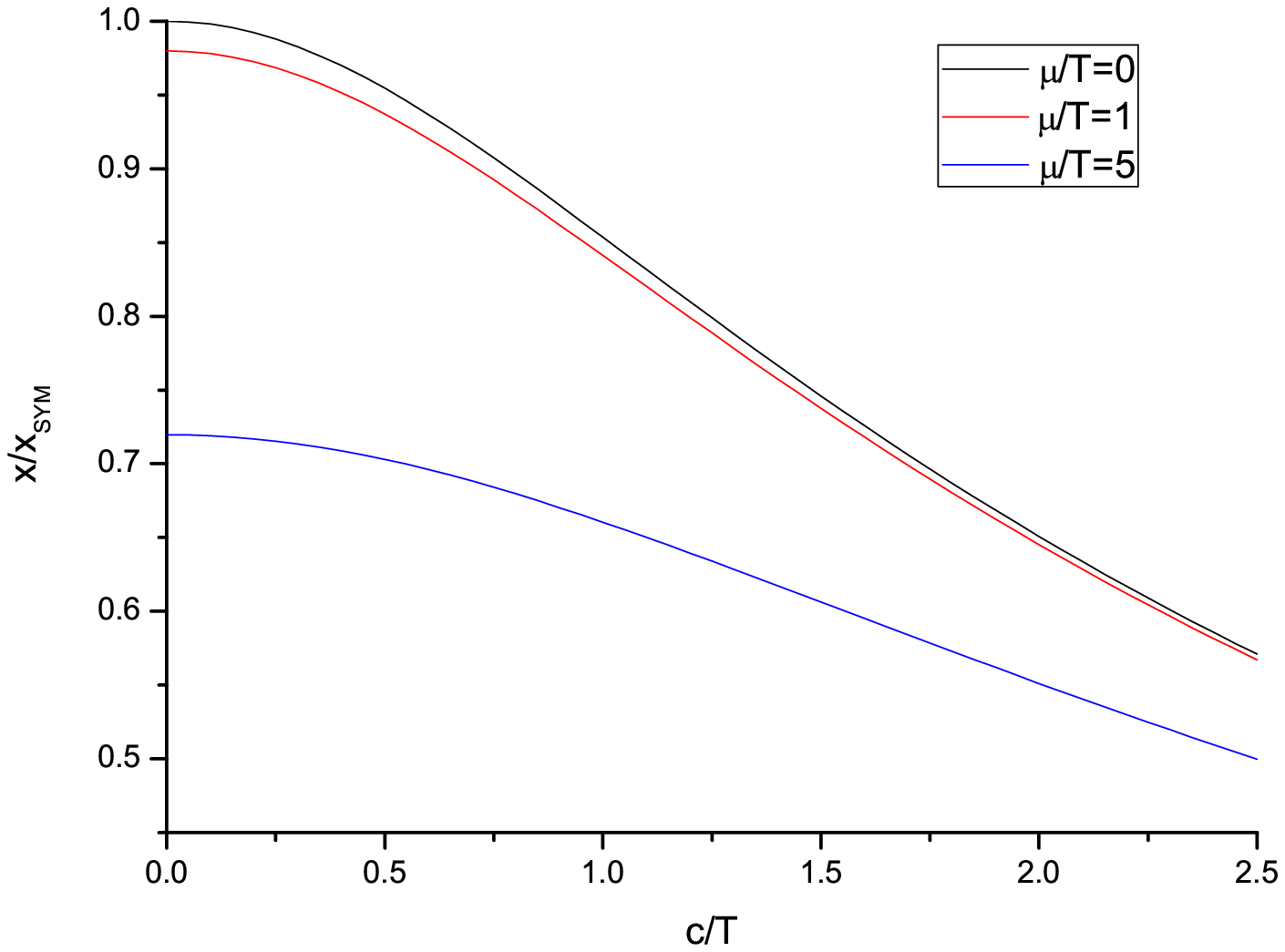}
\caption{Left: $x/x_{SYM}$ versus $\mu/T$. Right: $x/x_{SYM}$
versus $c/T$. Here we take $|\vec{q}|=0.99\omega$.}
\end{figure}

\subsection{Shooting string}
This subsection is devoted to the analysis of the light quarks
energy loss using shooting string. According to \cite{ss1,ss2},
one considers a particular type of classical string motion: the
string endpoint is close to horizon initially and move towards the
boundary, carrying some energy and momentum which are gradually
bled off into the rest of the string during its rise, hence, this
motion is called finite-endpoint-momentum shooting string, or
shooting string for short.

Next, we will follow the approach in \cite{ss1,ss2} to study the
light quark energy loss within shooting string for the background
metric (\ref{metric}). The instantaneous energy loss of light
quarks takes the form
\begin{equation}
\frac{dE}{dx}=-\frac{|L|}{2\pi\alpha^\prime}\frac{1}{z^2},\label{de0}
\end{equation}
where $L$ represents the null geodesics that the endpoint follows.
From the above equation, it appears that small $z$ (meaning the
endpoint starts near the boundary) will result in large energy
loss, indicating the jets will be quenched quickly and cannot be
seen. So to hedge against this, one assumes the strings start
close to the horizon.

As before, one supposes the quark moving along the $x_3$
direction. Given that, the energy and momentum of the quark become
\begin{equation}
E=-\frac{h(z)}{\eta}\frac{f(z)}{z^2},
\end{equation}
and
\begin{equation}
p_{x_3}=\frac{h(z)}{\eta z^2}\frac{dx}{dt},
\end{equation}
with $\eta$ the auxiliary field. Then the null geodesics reads
\begin{equation}
L=\frac{E}{p_{x_3}}=-f(z)\frac{dt}{dx}.
\end{equation}

The finite momentum endpoints will move along $ds^2=0$, which
gives
\begin{equation}
(\frac{dx}{dz})^2=\frac{1}{L^2-f(z)}. \label{dx}
\end{equation}

The denominator of (\ref{dx}) will vanish at $z=z_*$, yielding
\begin{equation}
L=-\sqrt{f(z_*)}. \label{nu0}
\end{equation}

So the null geodesics equation becomes
\begin{equation}
\frac{dx}{dz}=\frac{1}{\sqrt{f(z_*)-f(z)}}.\label{nul}
\end{equation}

Finally, using (\ref{metric}), (\ref{de0}) and (\ref{nu0}), one
obtains the energy loss for the SW$_{T,\mu}$ model as
\begin{equation}
\frac{dE}{dx}=-\frac{e^{c^2z^2}}{2\pi\alpha^\prime}\frac{\sqrt{1-(1+\frac{\mu^2z_h^2}{3})\frac{z_*^4}{z_h^4}
+\frac{\mu^2z_*^6}{3z_h^4}}}{z^2},\label{de1}
\end{equation}
note that for $\mu=c=0$, the above equation recovers the result of
SYM \cite{ss1,ss2}.

Before going further, one needs to solve the null geodesics
equation (\ref{nul}). However, it is difficult to solve it
analytically, but it is possible numerically. The procedures of
the numerical evaluation are as follows: First, sending
$z_*\rightarrow0$, then for a given value of $\mu/T$, one can
integrate (\ref{nul}) and invert to get $z(x)$. Next, substituting
$z(x)$ into (\ref{de1}) one can obtain $dE/dx$ as a function of
$c/T$, $\mu/T$ as well as $x$.

In fig.2, we compare the instantaneous energy loss of a light
quark moving in the SW$_{T,\mu}$ model with its counterpart in
$\mathcal N=4$ SYM plasma. The left panel corresponds to
$(dE/dx)/(dE/dx)_{SYM}$ versus $\mu/T$ with fixed $c/T$ while the
right one denotes $(dE/dx)/(dE/dx)_{SYM}$ versus $c/T$ with fixed
$\mu/T$, where $(dE/dx)_{SYM}$ represents the energy loss of SYM.
For both panels, one finds the energy loss in the SW$_{T,\mu}$
model is larger than that of SYM. In particular, increasing
$\mu/T$ or $c/T$ both increase $(dE/dx)/(dE/dx)_{SYM}$. Namely,
the inclusion of chemical potential and confining scale both
increase the energy loss, consistently with the analysis of the
stopping distance in the last subsection. The physical
significance of the results will be discussed in the next section.

\begin{figure}
\centering
\includegraphics[width=8cm]{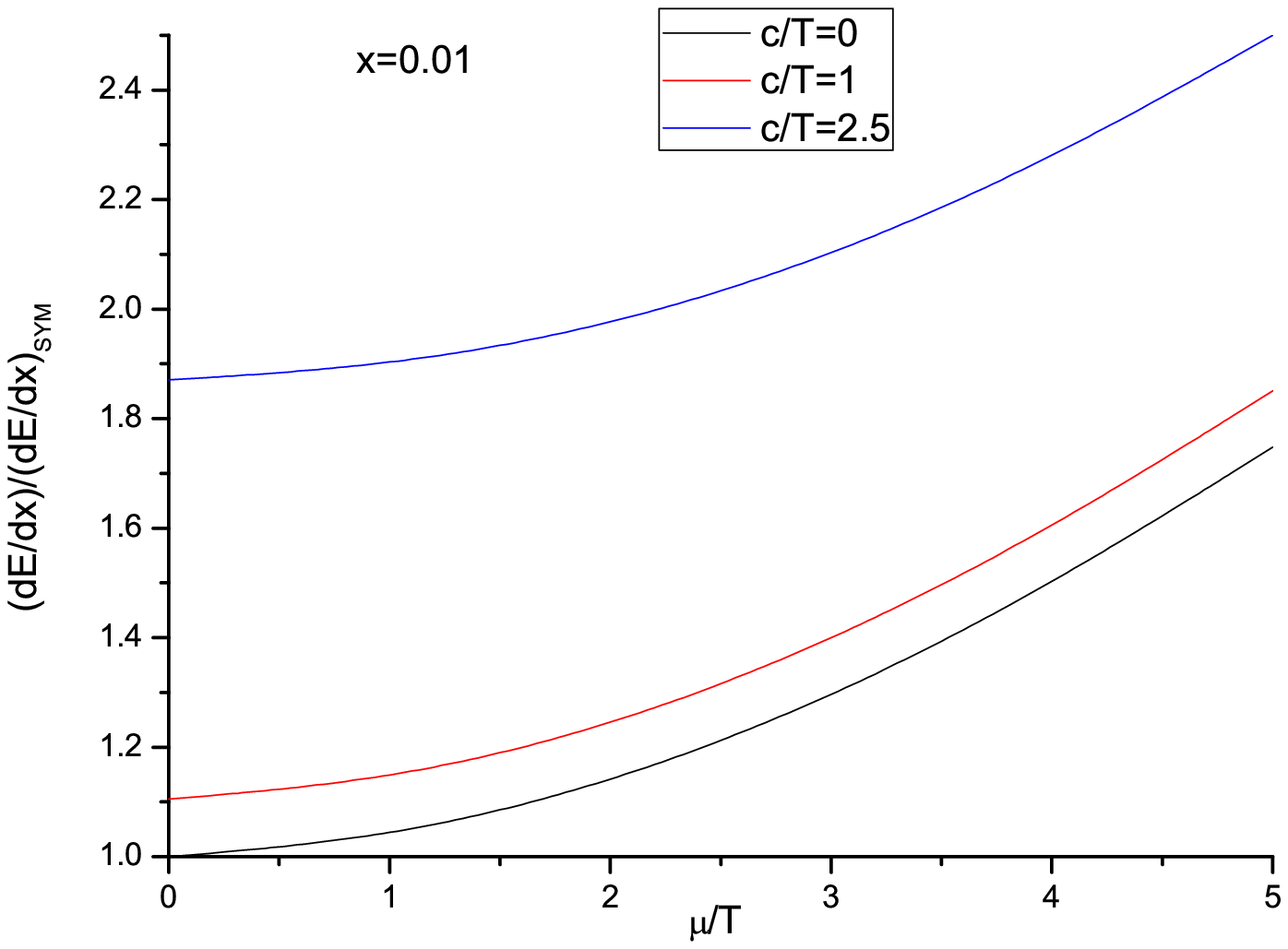}
\includegraphics[width=8cm]{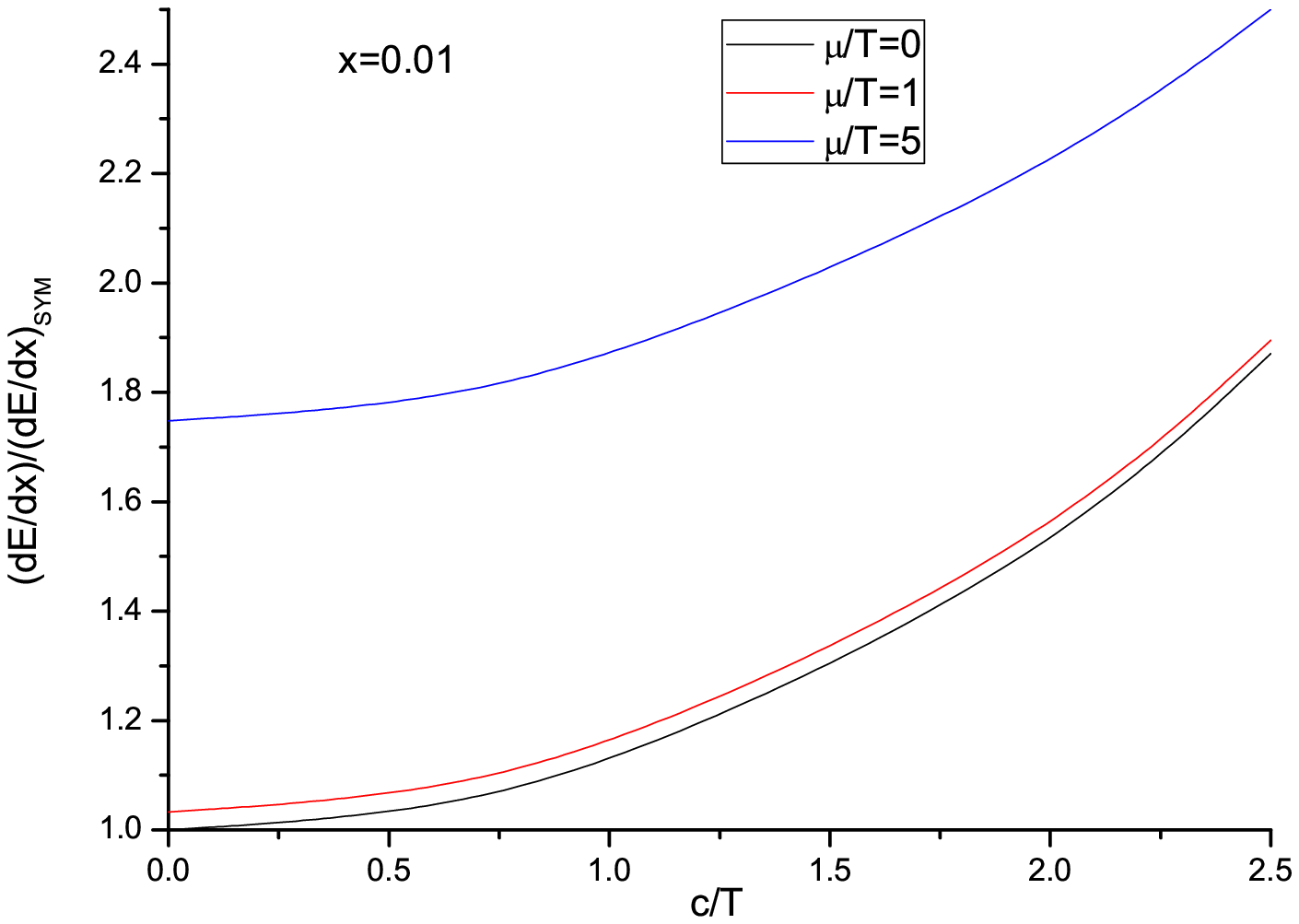}
\caption{Left: $(dE/dx)/(dE/dx)_{SYM}$ versus $\mu/T$. Right:
$(dE/dx)/(dE/dx)_{SYM}$ versus $c/T$. Here we take $x=0.01$.}
\end{figure}

\section{conclusion and discussion}
Jet quenching in high-energy heavy-ion collisions can be used to
probe properties of QGP. Studying the jet quenching in strongly
coupled nonconformal plasma with finite quark density may shed
qualitative insights into analogous questions in QCD. In this
paper, we investigated the energy loss of light quarks in the
SW$_{T,\mu}$ model using falling string and shooting string, in
turn. We discussed how chemical potential and confining scale
modify the energy loss for both cases, respectively. For the
former, we calculated the stopping distance of a massless particle
moving along the null geodesic and found the presence of chemical
potential and confining scale both decrease the stopping distance
thus increasing the energy loss. For the latter, we computed the
instantaneous energy loss based on the finite endpoint momentum
framework and found the same result: increase chemical potential
and confining scale both enhance the energy loss. These results
agree with those obtained by the drag force and jet quenching
parameter in the SW$_{T,\mu}$ model \cite{YH,zq0}. Taking all of
this together, one may draw a conclusion that the effects of
chemical potential and confining scale on the energy loss of heavy
quarks and light quarks are consistent.

We would like to make some comment on the possible significance of
our results to the current study on heavy-ion physics although
such comparison may be qualitative. Recent results of the
experiments at RHIC and LHC indicate that after collision, QGP
expands and the jet quenching decreases with a decrease in
temperature. Since chemical potential and confining scale both
have the effect of increasing the energy loss, one may infer that
including the two effects may lower the required temperature
region in order to observe the obtained RHIC or LHC values for jet
quenching.

However, it must be admitted that there are some limitations to
the present study. The major disadvantage is that the SW$_{T,\mu}$
model is not a bonafide solution of the classical equations of
motion. Considering jet quenching in some consistent models, e.g.
\cite{JP,AST,DL,DL1,SH,SH1,RRO} would be instructive. We hope to
report our progress in this regard in the near future.

\section{Acknowledgments}
This work is supported by Zhejiang Provincial Natural Science
Foundation of China Nos. LY19A050001, LY18A050002 and the NSFC
under Grant No. 11947410.

%%%%%%%%%%%%%%%%%%%%%%%%%%%%%%%%%%%%%%%%

\end{document}